\documentclass[12pt]{article}

\pdfoutput=1

\usepackage[english]{babel}
\usepackage{fancyhdr}
\usepackage{amsmath}
\usepackage{amssymb}
\usepackage{amsfonts}
\usepackage{psfrag}
\usepackage[applemac]{inputenc}
\usepackage{graphicx,wrapfig}
\usepackage{pstricks}

\usepackage{slashed}            
\usepackage{subfig}             
\usepackage{xspace}				
\usepackage[sort&compress,square,comma,numbers]{natbib}
\usepackage{float}
\usepackage{dcolumn}
\usepackage{bm}
\usepackage{epstopdf}

\usepackage[unicode=true,pdfusetitle,
 bookmarks=true,bookmarksnumbered=false,bookmarksopen=false,
 breaklinks=false,pdfborder={0 0 1},backref=false,colorlinks=true]
 {hyperref}

\DeclareFontFamily{OT1}{pzc}{}
\DeclareFontShape{OT1}{pzc}{m}{it}{<-> s * [1.100] pzcmi7t}{}
\DeclareMathAlphabet{\mathpzc}{OT1}{pzc}{m}{it}

\newcommand{\Tr}{\mathop{\rm Tr}}

\newcommand{\beq}{\begin{eqnarray}}
\newcommand{\eeq}{\end{eqnarray}}
\newcommand{\bea}{\begin{eqnarray}}
\newcommand{\eea}{\end{eqnarray}}
\newcommand{\bag}{\begin{align}}
\newcommand{\eag}{\end{align}}

\newcommand{\ie}{$\textnormal{i.e.}$ }

\newcommand{\MeV}{\,\mathrm{MeV}}
\newcommand{\GeV}{\,\mathrm{GeV}}
\newcommand{\TeV}{\,\mathrm{TeV}}

\newcommand{\fb}{\,\mathrm{fb}}

\newcommand{\eq}[1]{Eq.~(\ref{#1})}

\newcommand{\BR}{\,\mathrm{BR}}

\begin{document}

\baselineskip=18pt

\setcounter{footnote}{0}
\setcounter{figure}{0}
\setcounter{table}{0}


\begin{titlepage}

\hfill{CERN-PH-TH-2015-307}

\hfill{Saclay-t15/214}
\vspace{1cm}

\begin{center}
  \begin{LARGE}
    \begin{bf}
Goldstones in Diphotons
  \end{bf}
  \end{LARGE}
\end{center}
\vspace{0.1cm}
\begin{center}
\begin{large}
{\bf Brando Bellazzini$^{a,b}$, Roberto Franceschini$^c$,  Filippo Sala$^d$,  Javi Serra$^c$}
\end{large}
  \vspace{0.2cm}
  \begin{it}

\begin{small}
$^{(a)}$Institut de Physique Th\'eorique, Universit\'e Paris Saclay, CEA, CNRS, F-91191 Gif-sur-Yvette, France
\vspace{0.2cm}\\
$^{(b)}$Dipartimento di Fisica e Astronomia, Universit\'a di Padova, 
Via Marzolo 8, I-35131 Padova, Italy
\vspace{0.2cm}\\
$^{(c)}$CERN, Theory Division, Geneva, Switzerland
\vspace{0.2cm}\\
$^{(d)}$ LPTHE, CNRS, UMR 7589, 4 Place Jussieu, F-75252, Paris, France
 \vspace{0.1cm}

\end{small}

\end{it}
\vspace{.5cm}

\end{center}

\begin{abstract}
\medskip
\noindent
We study the conditions for a new 
scalar resonance to be observed first in diphotons at the LHC Run-2. 
We focus on scenarios where the scalar arises either from an internal or spacetime symmetry broken spontaneously, for which the mass is naturally below the cutoff and the low-energy interactions are fixed by the couplings to the broken currents, UV anomalies, and selection rules. 
We discuss the recent excess in diphoton resonance searches observed by ATLAS and CMS at 750 GeV, and explore its compatibility with other searches at Run-1 and its interpretation as Goldstone bosons in supersymmetry and composite Higgs models. 
We show that two candidates naturally emerge: a Goldstone boson from an internal symmetry with electromagnetic 
anomalies, and the scalar partner of the Goldstone of supersymmetry breaking: the sgoldstino. 
The dilaton from conformal symmetry breaking is instead disfavoured by present data, in its minimal natural realization. 
\end{abstract}

\bigskip

\end{titlepage}

\section{Introduction} \label{intro}

After the discovery of a light Higgs at the LHC Run-1, experimental efforts have shifted towards the search for new scalars beyond the SM. 
Evidence of another spin-0 resonance would be indicative of a natural solution of the electroweak hierarchy problem, as opposed to anthropic arguments, given that elementary scalars are generically unnatural in quantum field theory. 
As a matter of fact, extra scalars are common in both the two prototype solutions of the hierarchy problem, supersymmetry (SUSY) and composite Higgs models. 
Their experimental signatures at colliders are however model dependent, since they depend on the different selection rules of the specific realizations. 
An interesting question which we wish to address in this work is what are the conditions for the leading signal to be found in the diphoton channel $pp \to \Phi \to \gamma \gamma$, consistently with null resonance searches in other experimental analyses. 
We are motivated by the anomaly~\cite{15dec15,ATLAS:2015dxe,CMS:2015dxe} observed at the LHC by the ATLAS and CMS collaborations, which we take as a benchmark for a new scalar resonance $\Phi$ glimpsed for the first time in the diphoton channel. 

The anomaly is characterized by the signal strength, $\mu_{13}^{\gamma \gamma} = \sigma_{13\TeV} \times \BR_{\gamma\gamma}$, mass and  total width shown in Table~\ref{tab1}.
The difference between the observed and expected number of events in the bins of the excess is presently around 10. Therefore the statistical uncertainty on the  excess signal strength is, at best, a $\mathrm{few} \times 10\%$. 
Given the small number of signal events, it is premature to rely on any indication of a large resonance width. 
In fact, ATLAS reported a large local significance for both the narrow width and large width line-shapes used to characterize the excess.
We therefore treat the preferred width quoted by ATLAS \cite{ATLAS:2015dxe} as an upper bound.
Despite the small number of excess events, the relatively small amount of expected background events, and the bounds from Run-1 searches in the same final state, make it reasonable to speculate that such a signal is indeed the first glimpse of BSM physics at the LHC.

\begin{table}[htp]
\bigskip
\begin{center}
\begin{tabular}{|c|c|c|}
\hline
  $\mu_{13}^{\gamma \gamma}$ (fb)	&$m$ (GeV)	&  $\Gamma$ (GeV)   \\
\hline 
      7 & 750 & $\leq 45$   \\
\hline
\end{tabular}
\caption{\label{tab1} Signal strength in diphotons, mass and width suggested by the ATLAS and CMS analyses.}
\end{center}
\end{table}

A priori either a spin-0 or spin-2 boson, neutral under electromagnetism and colour, could reproduce such an excess over the SM expectation. 
Although with a poor significance, the observed width-over-mass ratio $\Gamma/m \approx 6\%$ 
would indicate that some of the effective couplings $g_\Phi$ of $\Phi$ to lighter states, either in the SM or beyond, should be somewhat sizeable, 
\begin{equation}
\frac{ \Gamma_\Phi}{m_\Phi} \sim \left(\frac{g_\Phi}{4\pi} \right)^2 \pi N \sim 6 \% \left( \frac{g_\Phi}{1} \right)^2 \left( \frac{N}{3} \right) \, , 
\label{width}
\end{equation} 
where $N$ is the multiplicity factor of the final state, including polarizations. 
We are defining the effective coupling $g_\Phi$ in a loose sense through the decay amplitude $\mathcal{M}_{1\rightarrow 2} \sim g_\Phi m_\Phi$. 
For instance, the effective coupling of $\Phi$ in a derivative interaction such as $c_H \Phi /f |D_\mu H|^2$ is set by the energy squared as $g_\Phi \sim c_H E^2/(f m_\Phi) = c_H m_\Phi/ f$.\footnote{Notice that the conclusion extracted from \eq{width} relies on the assumption that a single resonance contributes to the excess.}
The observed signal strength at LHC Run-2, along with the exclusion bound on diphotons at LHC Run-1, are reported in Table~\ref{tab2}. They place a lower bound on the ratio of production cross sections $\sigma_{13\TeV}/\sigma_{8\TeV}\gtrsim 4.0$, that indicates that the dominant production mode of the new resonance could be gluon fusion $gg \rightarrow \Phi$, for which $(\sigma_{13\TeV}/\sigma_{8\TeV})_{gg \rightarrow \Phi}\approx4.7$~\cite{Martin:2009iq}.
Hereafter we will consider resonances produced in this channel (and comment shortly on photon fusion). We will also restrict the discussion to the case of $\Phi$ being a spin-0 resonance, and assume invariance under CP (motivated by the nonobservation of CP violation beyond the SM);  we refer in the following to $\Phi=\sigma$ and $\Phi=\eta$ for a CP-even and CP-odd scalar boson, respectively. 

In Table~{\ref{tab2} we also show the lower bounds on the ratios of decay rates, $r^{\gamma}_{XY}\equiv \Gamma_{\gamma\gamma}/\Gamma_{XY}$, set by null searches at LHC Run-1 in the various channels other than the decay into diphotons. 
In deriving those constraints, we assumed that the production cross section scales with the $gg$ parton luminosity, using the MSTW2008 parton distribution functions \cite{Martin:2009iq} for definiteness. 
Strong bounds come from 8 TeV resonance searches in leptons, $\tau$ pairs, and dijets. 
However, from a theoretical point of view, these constraints are not very severe or difficult to evade, as chiral symmetry controls the size of the Yukawa couplings of $\Phi$ through the fermion masses.
\begin{table}[!tp]
\begin{center}
{\scriptsize \footnotesize 
\begin{tabular}{|c|c|ccccccccc|}
\hline
	& $ {\sigma_{13\TeV} \over \sigma_{8\TeV}} $  &  $10^2 \times$&$r^{\gamma}_{WW}$  &  $r^{\gamma}_{ZZ}$  &  $r^{\gamma}_{Z\gamma}$  & $r^{\gamma}_{hh}$   &  $r^{\gamma}_{t\bar{t}}$    & $r^{\gamma}_{\tau\bar{\tau}}$ & $r^{\gamma}_{\ell\bar{\ell}}$ & $r^{\gamma}_{gg}$   \\
\hline 
 ATLAS &    2.9~\cite{Aad:2015mna}$^* $  & &  3.0~\cite{Aad:2015agg}  &    13~\cite{Aad:2015kna}$^* $  & 19~\cite{Aad:2014fha}$^* $ &  4.1~\cite{Aad:2015xja}$^*$ &  0.22~\cite{Aad:2015fna}  &   15~\cite{Aad:2015osa} &  124~\cite{Aad:2014cka}$^*$   & $0.14$~\cite{Aad:2014aqa}  \\
 \hline
CMS &      4.0~\cite{Khachatryan:2015qba}  & & $0.5$~\cite{Khachatryan:2015cwa}  &   4.6~\cite{Khachatryan:2015cwa} & -- &  2.8~\cite{Khachatryan:2015yea}$^* $  &  $0.33$~\cite{Chatrchyan:2013lca} &    7.4~\cite{CMS:2015mca}  &  114~\cite{Khachatryan:2014fba}$^*$   & 0.083~\cite{CMS:2015neg}$^*$  \\
   \hline
\end{tabular}
}
\caption{\label{tab2} Experimental lower bounds on the ratio of production cross sections at 13 and 8 TeV, and on $10^2 \times r^{\gamma}_{XY}$ from the $8$ TeV run of LHC, where $r^{\gamma}_{XY}\equiv~\Gamma_{\gamma\gamma}/\Gamma_{XY}$. We dub with a star * the bounds where the width assumed in the corresponding experimental analysis is smaller than 45~GeV.}
\end{center}
\end{table}%

With this theoretical prior, we assume the following structure for the effective phenomenological lagrangian of $\Phi$:
\begin{equation}
\mathcal{L}= \mathcal{L}^{SM}+ \frac{1}{2}(\partial_\mu\Phi)^2 -\frac{1}{2}m_\Phi^2 \Phi^2 +\mathcal{L}_I^{\Phi} \ , \qquad \Phi=\sigma \, , \eta \, ,
\end{equation}
where
\begin{align}
 \label{ffsigma}
\mathcal{L}_I^{\Phi=\sigma} = & \frac{\sigma}{f} \left( c_h \partial_\mu h  \partial^\mu h + c_Z m_Z^2 Z_{\mu}Z^\mu  + 2 c_W m_W^2 W_{\mu}W^\mu \right)  \\
\nonumber
& -  \frac{\sigma}{f} \left( c_t m_t \bar{t}t + c_b m_b \bar{b} b+ c_\tau m_\tau \bar{\tau}\tau \right) \\ 
\nonumber
& + \frac{\sigma}{f} \left(c_{gg} \frac{\alpha_3}{8\pi} G_{\mu\nu}^2 + c_{\gamma\gamma} \frac{\alpha_e}{8\pi} F_{\mu\nu}^2 \right) \\ \nonumber 
&+ \frac{\sigma}{f} \left( c_{WW} \frac{\alpha_2}{4\pi} W^+_{\mu\nu} W^{-\,\mu\nu} + c_{ZZ} \frac{\alpha_2 \cos^2\theta_W}{8\pi} Z_{\mu\nu}^2 + c_{Z\gamma}\frac{\alpha_e}{4\pi\tan\theta_W} Z_{\mu\nu}F^{\mu\nu} \right) \\
\nonumber
& - \frac{\sigma}{f} c_3 m_h^2 h^2 \,,
\end{align}
and
\begin{align}
\label{Leffeta}
\mathcal{L}^{\Phi=\eta}_I = & -  i\frac{\eta}{f} \left( C_t m_t \bar{t}\gamma^5 t + C_b m_b \bar{b}\gamma^5 b+ C_\tau m_\tau \bar{\tau}\gamma^5\tau \right) \\
\nonumber
& - \frac{\eta}{f} \left( C_{gg} \frac{\alpha_3}{8\pi} G^a_{\mu\nu} \tilde G^{a\,\mu\nu} + C_{\gamma\gamma} \frac{\alpha_e}{8\pi} F_{\mu\nu} \tilde F_{\mu\nu} \right) \label{ffeta} \\
\nonumber
& - \frac{\eta}{f} \left( C_{WW} \frac{\alpha_2}{4\pi} W^+_{\mu\nu} \tilde W^{-\, \mu\nu} +C_{ZZ}\frac{\alpha_2\cos^2\theta_W}{8\pi} Z_{\mu\nu} \tilde Z^{\mu\nu} +C_{Z\gamma}\frac{\alpha_e}{4\pi\tan\theta_W} Z_{\mu\nu} \tilde F^{\mu\nu} \right) \,,
\end{align}
where $\tilde F^{\mu \nu} \equiv \epsilon^{\mu\nu\rho\sigma} F_{\rho\sigma}$ for any of the field strengths, and $\theta_W$ is the weak angle. 
We did not include terms with more than one $\Phi$-leg, given they are irrelevant for the phenomenology we discuss in this paper. 
In full analogy with the electroweak vacuum expectation value $v\simeq 246 \GeV$ or the pion decay constant in QCD $f_\pi$, the scale $f$ appearing in the effective lagrangians above does not represent in general a physical mass threshold. 
Instead the mass of the heavy BSM states associated with the scale $f$ are controlled by that scale times a coupling $m_* = g_* f$, like $m_W=g v/2$ or $m_\rho\approx g_\rho f_\pi$. 
We did not introduce flavor indices, and we will be assuming no flavor violating couplings for $\Phi$. 
Notice that we parametrize the couplings to the gauge field strengths following the generic expectation in weakly coupled theories of a loop suppression $\alpha/4\pi$. 
However this expectation is challenged by the signal excess, which neglecting $K$ factors we find to be well approximated by
\beq
\label{sigmawidthnum}
\mu_{13, \, \sigma}^{\gamma \gamma} \simeq 7 \fb \left( \frac{|c_{gg}^{eff}| |c_{\gamma \gamma}^{eff}|}{500} \right)^2 \left( \frac{500 \GeV}{f} \right)^4 \left( \frac{45 \GeV}{\Gamma_\sigma} \right) \, ,
\eeq
\beq
\label{etawidthnum}
\mu_{13, \, \eta}^{\gamma \gamma} =\, \mu_{13, \, \sigma}^{\gamma \gamma} (c_{gg,\gamma\gamma}^{eff} \to 2 \, C_{gg,\gamma\gamma}^{eff}) \ ,
\eeq
where  $|c_{gg,\gamma\gamma}^{eff}|$ and $|C_{gg,\gamma\gamma}^{eff}|$ parametrize the effective coupling to $gg$ and $\gamma \gamma$ from the contact terms $c_{gg,\gamma\gamma}$, and $C_{gg,\gamma\gamma}$ respectively, as well as from loops of the SM fermions and $W$'s. See Eqs.~(\ref{Ceffeta}) and (\ref{Ceffdilaton}) for the impact of the loops in $|C^{eff}_{gg,\gamma\gamma}|$ and $|c_{gg,\gamma\gamma}^{eff}|$ respectively. 
We advance that the requirement of large $|c_{\gamma\gamma}^{eff}|$ or $|C_{\gamma\gamma}^{eff}|$, combined with the experimental constrains from $t \bar t$ and $VV$, $V=\{W,\, Z\}$, discussed below, implies that if no new physics contributions to the effective couplings of $\Phi$ to photons are present ($c_{\gamma\gamma} =C_{\gamma\gamma} = 0$), the excess cannot be reproduced.\\

Let us remark at this point our interpretation of the resonance $\Phi$ as a pseudo-Goldstone boson (pGB) of some kind, to be expanded in the sections below.
The required sizable couplings of a generic scalar $\Phi$ to photons (gluons) can arise from loops of multiple electrically (colour) charged states, $X$, coupled to $\Phi$.
Nevertheless, on fairly general grounds one expects these states to contribute to the mass of $\Phi$ at the same loop order,
\beq
(\Delta m^{2}_\Phi)_X \sim N_X \frac{y_X^2}{16 \pi^2} m_*^2 \, , 
\eeq
with $N_X$ their multiplicity and $y_X$ their coupling to $\Phi$. 
This suggests that in the absence of a protection mechanism or tuning, $\Phi$ would not be separated from the cutoff $m_*$ of our effective field theory, possibly compromising calculability and predictivity. 
This is just the familiar hierarchy problem that plagues any theory with unprotected scalars.
This argument holds for weakly coupled theories as well as for strongly coupled ones where $\Phi$ arises as a composite state of the underlying dynamics. 
For the latter, even though $X$ particles might not be required to generate large couplings to photons (gluons), that is $c_{\gamma\gamma} (c_{gg}) \gg 1$, one expects $m_\Phi \ll m_*$ only by accident.
Because of these reasons, an (approximate) shift-symmetry acting on $\Phi$ will be our rationale for its lightness compared to $m_*$. 
The same selection rules control also the $\Phi$ couplings that do not respect the shift symmetry (this will become apparent in the sections that follow).
\\

Since we consider production via gluon fusion, a non-vanishing decay rate to gluons is constrained by 8 TeV resonance searches in dijets,
\beq
\label{jjbound}
r^\gamma_{gg} \simeq \frac{1}{8} \left( \frac{\alpha_e}{\alpha_3} \right)^2 \left( \frac{|c_{\gamma \gamma}^{eff}|}{|c_{gg}^{eff}|} \right)^2 \, ,
\eeq
for both the scalar and pseudo-scalar ($c_{gg,\gamma\gamma}^{eff} \to C_{gg,\gamma\gamma}^{eff}$) candidates.

Furthermore, given the interactions of the scalar $\sigma$, we can identify the most stringent constraints as those coming from $pp \to \Phi \rightarrow VV, hh$ where $V=\{W,\, Z\}$.
The corresponding decay widths scale with $m_\sigma^3$, and the theoretical prediction for the ratio of decay rates reads
\begin{equation}
\Phi=\sigma: \quad r^\gamma_{VV}\simeq n_V \left(\frac{\alpha_e}{4\pi}\right)^2 \left(\frac{c_{\gamma\gamma}}{c_V}\right)^2 \,, \qquad 
r^\gamma_{hh}\simeq 2 \left(\frac{\alpha_e}{4\pi}\right)^2 \left(\frac{c_{\gamma\gamma}}{c_h}\right)^2\,.
\label{VVbound}
\end{equation}
where $n_{Z} = 1$ and $n_{W} = 2$. 
Compatibility with the corresponding bounds shown in Table~\ref{tab2} requires very large $c_{\gamma\gamma}/c_{h,V}=O(16\pi^2)$. 
In these general estimates, just for simplicity of the exposition, we neglected $O(m_{V,h}/m_\sigma)$ and the top loop contribution to $\gamma\gamma$ (notice that the contribution to $\Gamma_{hh}$ from $c_3$ is generically suppressed by $m_h^2/m_\sigma^2$). 
We point out also that in models with custodial symmetry $c_h = c_W = c_Z$, and in that case the strongest constraint comes from $ZZ$ resonance searches at Run-1.
Such large values of $c_{\gamma\gamma}/c_{h,V}$ can be obtained by suppressing $c_{h,V}$ and/or boosting $c_{\gamma\gamma}$, as it may happen when the couplings to $VV$, $hh$ and $\gamma\gamma$ are generated at the same loop order, \ie either all at tree- or loop-level. In Section~\ref{sectSgold} we discuss such a scenario where $\sigma$ is the sgoldstino, for which all the couplings can arise at tree-level, resulting in $c_{\gamma\gamma}/c_{h,V} = O(16\pi^2)$. 
Alternatively, it may happen that while the coupling to photons is formally at one-loop and $c_{h,V}$ is tree-level, the former is endowed with an anomalously large factor like in the dilaton model we present in Section~\ref{sectDil}, where $c_{\gamma\gamma}$ counts new degrees of freedom charged under electromagnetism. 
  
The CP-odd scalar $\eta$ instead has a built-in selection rule, CP invariance, that forbids the tree-level coupling to the longitudinal $V$, and $h$, so that the bounds from $r^\gamma_{VV,hh}$ are satisfied more easily. 
Instead for the pseudo-scalar the bounds from $t\bar{t}$ can become relevant, as $C_t$ respects CP. 
However, since the partial decay width into tops scales with $m_\eta$ (to be contrasted with the one in diphotons that goes like $m_\eta^3$), and the bound on $r^{\gamma}_{t \bar t}$ is milder, a slightly smaller coupling $C_t$ or an enhanced anomaly coefficient $C_{\gamma\gamma}=O(\mathrm{few})$ are enough to satisfy the experimental constraints: 
\begin{equation}
\label{etattbarbound}
\Phi=\eta:\quad  r^\gamma_{t\bar{t}} \simeq \frac{2}{3}\left(\frac{\alpha_e}{4\pi}\right)^2 \left(\frac{m_\eta}{m_t}\right)^2 \left(\frac{|C_{\gamma\gamma}^{eff}|}{C_t} \right)^2 \, , \quad 
\frac{|C_{\gamma\gamma}^{eff}|}{C_t} = \left|\dfrac{C_{\gamma\gamma}}{C_t} + \frac{2}{3} \,A_{1/2}(x_t) \right| \, ,
\end{equation}
where $A_{1/2}(x_t)$, $x_t = 4m_t^2/m_\eta^2$, is the contribution of the top loop to the effective $\gamma \gamma$ coupling.
In the next section we discuss in detail this natural case in the context of a pGB emerging from an anomalous $U(1)$ that is broken spontaneously. 

In theories where the electroweak $SU(2)_L \times U(1)_Y$ symmetry is linearly realized above the scale $f > v$, such as the ones we consider in this paper, a non-trivial constraint arises from $Z\gamma$ or $ZZ$ resonance searches (for the latter even if $c_Z = 0$). 
This is clear once the $\gamma \gamma$ coupling for $\Phi$ is written in a manifestly symmetric form
\beq
\mathcal{L}_I^{\Phi=\eta} \supset -\frac{\eta}{f} \left( C_W \frac{\alpha_2}{8\pi} W^{i}_{\mu\nu} \tilde W^{i \, \mu \nu} + C_B \frac{\alpha_1}{8\pi} B_{\mu\nu} \tilde B^{\mu\nu} \right) \, ,
\eeq
which implies the relations
\begin{equation}
C_{\gamma\gamma}=C_W+C_B\,,\qquad C_{WW}=C_W \,,\qquad C_{ZZ}=C_W+\tan^4\theta_W C_B\,, \qquad C_{Z\gamma}=C_W-\tan^2\theta_W C_B \, ,
\label{looprelations}
\end{equation}
and equivalently for $\Phi = \sigma$.\\

Finally, let us note that qualitatively, most of the phenomenological features discussed above are fairly independent of the benchmark values $m_\Phi \approx 750 \GeV$ and $\mu_{13}^{\gamma \gamma} \approx 7 \fb$ associated with the diphoton excess. 
Given a resonance mass large enough to kinematically allow any of the decays considered above, again at a qualitative level the interplay of the dijet, $VV$ and $hh$ constraints with the diphoton signal is independent of the mass, since the widths into diboson final states scales with the same power of $m_\Phi$, see Eqs.~(\ref{jjbound}) and (\ref{VVbound}). 
The interplay with $t\bar{t}$ searches depends instead on $m_\Phi$, and this constraint becomes less important for heavier resonances. 
However, we will discuss both cases where fermionic decays play a role and where they do not, sections \ref{sectpGB} and \ref{sectSgold} respectively.
We expect these two cases to be representative of the possible phenomenology of a resonance whose leading signature is in diphotons.

\section{Pseudo-Goldstone boson}
\label{sectpGB}

From the insights of the previous section, we are led to consider the case where $\Phi=\eta$ is a SM singlet that emerges as the GB of a spontaneously broken internal $U(1)_{\eta}$ symmetry.
As we show below, we need such a $U(1)_\eta$ to be anomalous under $U(1)_{Y}$ (and/or $SU(2)_{L}$), while $SU(3)_C$ anomalies could also be present. 
This type of state could find its realization in composite Higgs models where the pattern of global symmetry breaking $G/H$ gives rise to the Higgs and the singlet as GBs,\footnote{
Generic composite Higgs models also predict the existence of extra scalar resonances, whose mass however is expected to be $m_* = g_* f$ and thus above the TeV, along with most of the strongly coupled resonances.}
and it is such that it admits a Wess-Zumino-Witten term (see e.g.~the discussion in \cite{Gripaios:2009pe}). 
Examples can be found in models based on the coset $SU(3) \times U(1)/SU(2) \times U(1)$ \cite{Contino:2003ve,Schmaltz:2004de,Hill:2007nz}, $SU(4)/Sp(4)$ \cite{Gripaios:2009pe,Katz:2005au,Galloway:2010bp,Barnard:2013zea,Serra:2015xfa,Cai:2015bss}, $SU(5)/SO(5)$ \cite{ArkaniHamed:2002qy,Vecchi:2013bja,Ferretti:2014qta}, $SU(3)^2/SU(3)$ \cite{ArkaniHamed:2002qx,Vecchi:2015fma} like in QCD (for which the pion has an electromagnetic anomaly), and their extensions to higher-rank groups (see \cite{Bellazzini:2014yua} for a review).%
\footnote{Notice that some of these cosets are not endowed with a custodial symmetry, like $SU(3) \times U(1)/SU(2) \times U(1)$ or $SU(3)^2/SU(3)$. Besides, in some of the models and depending on the specific realization, the singlet does not have an electromagnetic anomaly, like in \cite{Gripaios:2009pe}.}
In these examples, for which $SU(3)_C$ is factorized from the coset structure, there are no colour anomalies for the GBs. 
These could arise however in more involved models where the colour symmetry group is embedded in a non-trivial way. 
Besides, we show below that a colour anomaly, in contrast to the electromagnetic one, is not strictly required to reproduce the excess.

The linear coupling of $\eta$ are given by the effective lagrangian
\begin{align}
\mathcal{L}^\eta_I= &  -\frac{\eta}{f} \left( iC_t\frac{\sqrt{2}m_t}{v} \bar{q}_L \tilde H t_R +h.c. + \ldots \right) \label{yuketa} \\
& - \frac{\eta}{f} \left( C_{G} \frac{\alpha_3}{8\pi} G^a_{\mu\nu} \tilde G^{a \, \mu\nu}+C_W \frac{\alpha_2}{8\pi} W^{i}_{\mu\nu} \tilde W^{i \, \mu \nu} + C_B \frac{\alpha_1}{8\pi} B_{\mu\nu} \tilde B^{\mu\nu} \right) \ . \label{anometa}
\end{align}  
which reduces to the parametrization of Eq.~(\ref{Leffeta}) putting the Higgs to its VEV, keeping in mind the relations given in \eq{looprelations} and $C_{gg} = C_G$. 
The top Yukawa-like coupling $C_{t}$ in Eq.~(\ref{yuketa}) (and likewise the couplings to other SM fermions) breaks the shift symmetry $\eta\rightarrow \eta+c$, and so do the anomalous terms $C_{G,W,B}$ in Eq.~(\ref{anometa}). 
Therefore such terms generically contribute to the mass of the GB, see the discussion at the end of the section. 
Notice that another term can be added to the lagrangian of $\eta$, the standard coupling of the GB to the matter part of the broken current:
\begin{equation}
\label{etacurrent}
\frac{1}{f}\partial_\mu \eta J^{\mu}_{matter}=\frac{\partial_\mu\eta}{f} \left[c_{q_L}  \bar{q}_L \gamma^\mu q_L  + c_{t_R} \bar{t}_R \gamma^\mu t_R+\ldots +c_H \left(i H^\dagger (D^\mu H) +h.c. \right) \right]\,.
\end{equation}
The effects of such terms on the phenomenology of $\eta$ can be derived by the field redefinition
 \begin{align}
 \label{Qeta}
 q_L \rightarrow  e^{i c_{q_L} \eta/ f} q_L\,, \qquad 
t_R \rightarrow  e^{i c_{t_R} \eta/ f} t_R\,, \qquad 
H  \rightarrow    e^{i c_{H} \eta/ f} H\,,\qquad 
\ldots
 \end{align}
which changes the action as
  \begin{equation}
 S\rightarrow S+\int d^4x\, \frac{\eta}{f}\partial_\mu J^\mu\,,\qquad  \partial_\mu J^\mu=\partial_\mu J^\mu_{matter} +\mathcal{A}^{\eta ab}\frac{\alpha_V}{16\pi} V^a_{\mu\nu}  V^{b}_{\rho\sigma} \epsilon^{\mu\nu\rho\sigma}\,,
 \end{equation}
 where $\mathcal{A}^{\eta ab}=\Tr[T^a \{T^b, Q^{\eta}\}]$, and $Q^\eta=c_{q_L}, -c_{t_R},\ldots$ are the charges associated with the transformation (\ref{Qeta}) expressed for left-handed chiral fermions, and $\Tr[T^a T^b]=\delta^{ab}/2$ in the fundamental representation. 
The coupling to the current thus contributes to the Yukawas and to the anomalies in Eqs.~(\ref{yuketa}) and (\ref{anometa}).\footnote{The specific contributions are $\Delta C_t  = -c_{q_L} + c_{t_R}- c_{H}$, $\Delta C_{G} = -\frac{1}{2}(2 c_{q_L} -c_{t_R})$, $\Delta C_W = - \frac{3}{2} c_{q_L}$, and $\Delta C_B = -(\frac{1}{6}c_{q_L} -\frac{4}{3}c_{t_R})$.} \\

Invariance under CP  forbids the tree-level coupling to the longitudinal massive gauge bosons. Hence,  the pseudo-scalar GB couples to the gauge bosons, both massless and massive, at one-loop. 
The decay rates to photons and gluons are given by
 \begin{align}
 \label{Gammaeta}
 \Gamma_{\gamma\gamma}= \left( \dfrac{\alpha_e}{8\pi f}\right)^2 \dfrac{m_\eta^3}{\pi} \left|C^{eff}_{\gamma\gamma}\right|^2 \, , && \Gamma_{gg}=  8 \left(\dfrac{\alpha_3}{8 \pi f}\right)^2 \dfrac{m_\eta^3}{\pi} \left|C_{gg}^{eff}\right|^2\,,  \\
 \label{Ceffeta}
 C^{eff}_{\gamma\gamma}= C_{\gamma\gamma} + \frac{1}{2}C_t N_c^{(t)}Q_t^2 \,A_{1/2}(x_t)\,, && C^{eff}_{gg}= C_{gg} + \frac{1}{4}C_t  \,A_{1/2}(x_t) \,,
\end{align}
where $N_c^{(t)}Q_t^2=3\times\left(2/3\right)^2$, $x_t = 4m_t^2/m_\eta^2$, and $A_{1/2}(x)=2 x f(x)$ with $f(x)=-\frac{1}{4}\left(\log\frac{1+\sqrt{1-x}}{1-\sqrt{1-x}}-i\pi \right)^2$ for $x<1$ and $f(x)=\arcsin^2(1/\sqrt{x})$ for $x>1$. 
Notice we have only included the loop contribution from the top, which is linked to the decay rate of $\eta$ to top pairs,
\beq
\Gamma_{t \bar t} \simeq 3 \frac{C_t^2}{8 \pi} \frac{m_t^2 m_\eta}{f^2} \approx (40 \GeV) \left( \frac{C_t}{2} \right)^2 \left( \frac{500 \GeV}{f} \right)^2 \left( \frac{m_\eta}{750 \GeV} \right) \, .
\label{ttwidth}
\eeq
This decay channel dominates the total width of $\eta$ if the coupling to tops is sizable, and a large decay rate $\Gamma_\eta \approx 45 \GeV$ can be reproduced.
\begin{figure}[!t]
\centering
\includegraphics[width=8.5cm]{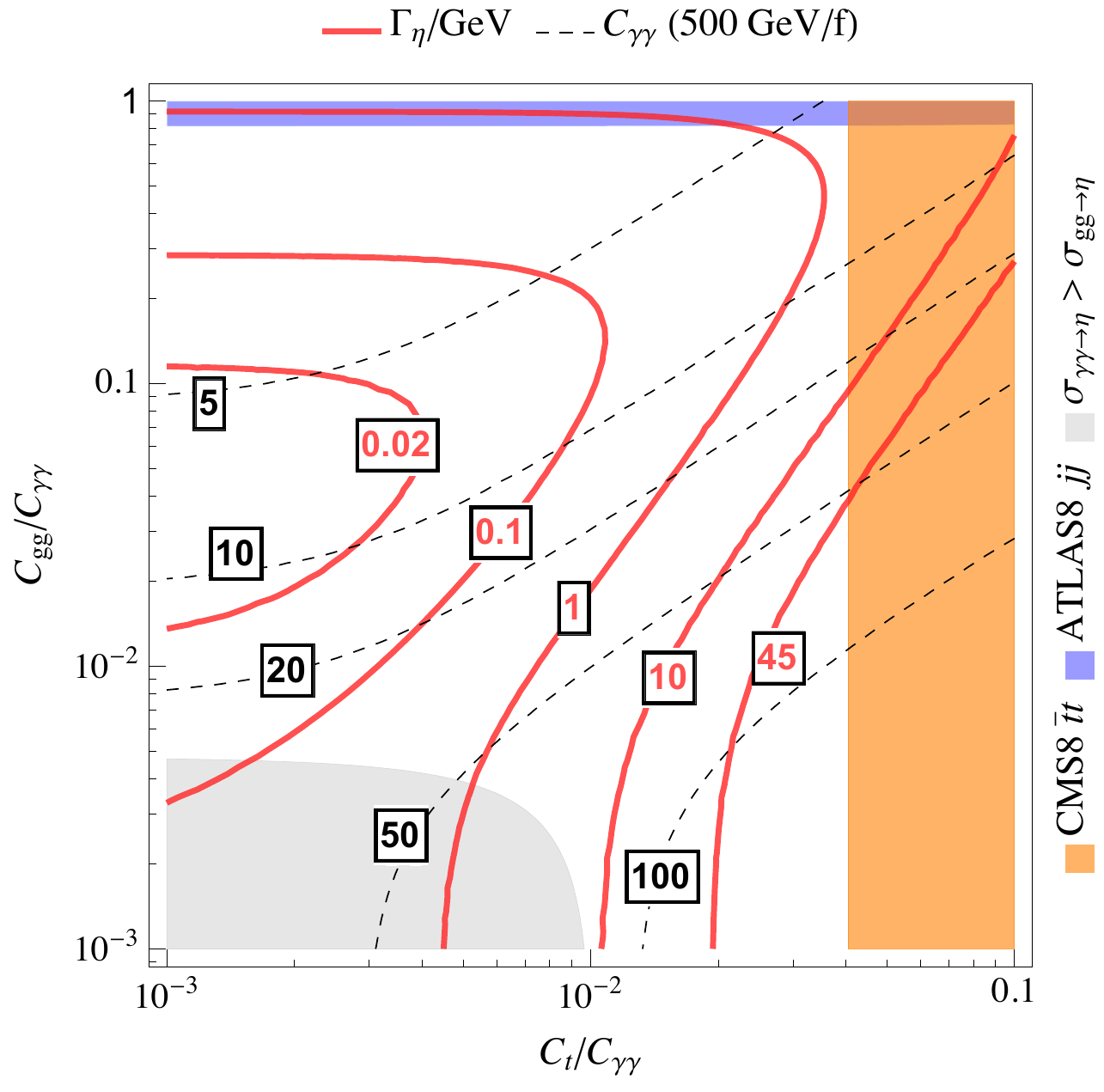}\hfill
\includegraphics[width=8.5 cm]{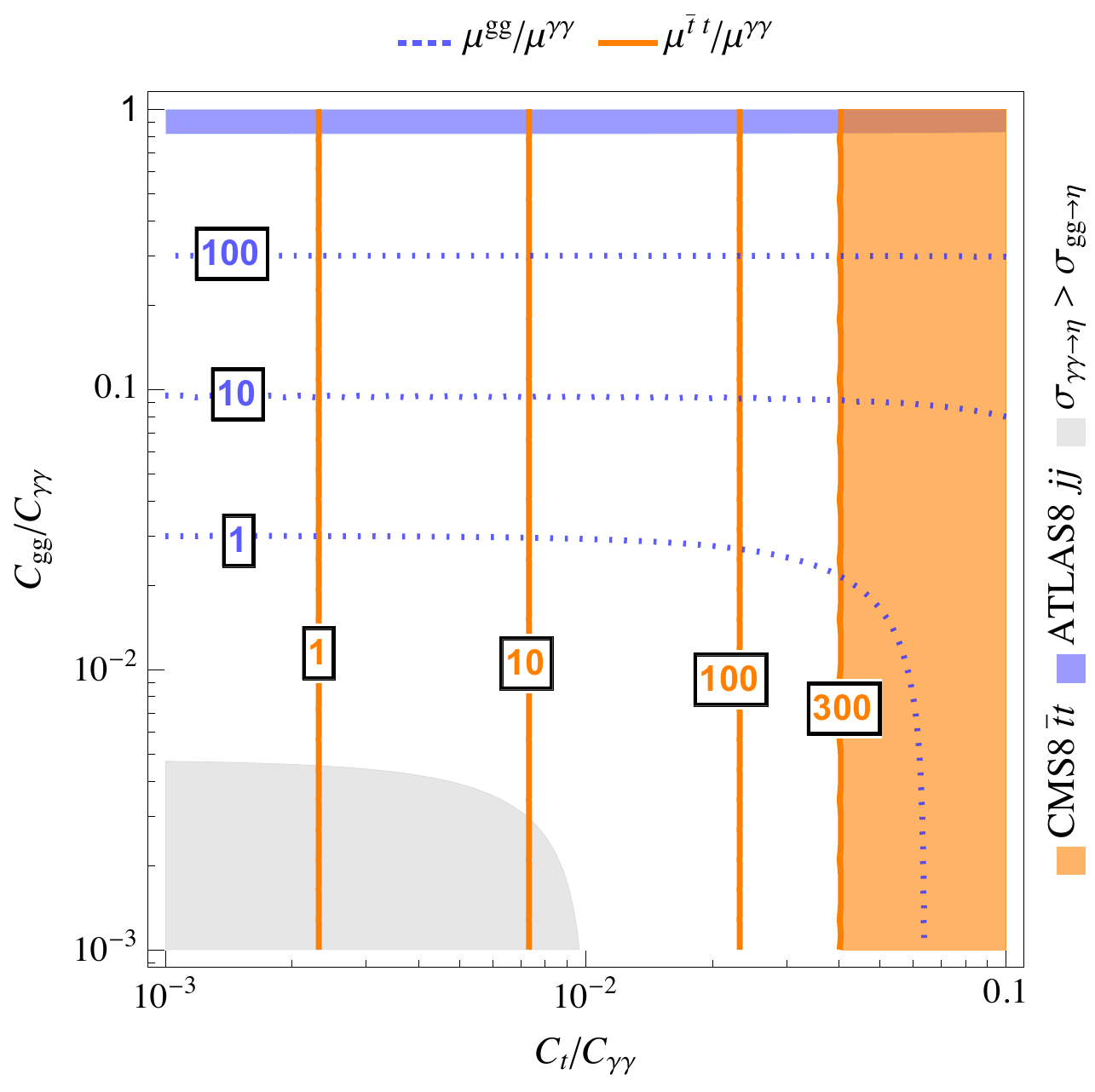}
\caption{Pseudo-GB parameter space, where we have fixed $m_\eta = 750 \GeV$ and $\mu^{\gamma\gamma}_{13} = 7 \fb$. Shaded regions are excluded by dijet (blue) and $t\bar{t}$ (orange) searches at 8 TeV. We have also shaded in grey the region where production via photon fusion dominates over gluon fusion (bottom-left corner). Left: the solid red line gives the expected width of the minimal GB model, the dashed black line the coupling to photons normalized to $f=500 \GeV$ $C_{\gamma \gamma} \, (500\GeV/f)$. Right: ratio of signals in $t \bar t$ (solid orange) and $gg$ (dotted blue) with respect to the signal in $\gamma\gamma$.}
\label{fig:goldstone}
\end{figure}
This requires a large UV coefficient $C_{\gamma \gamma} \gtrsim 50$ for $f = 500 \GeV$ in order to be compatible with $t\bar t$ searches \eq{etattbarbound}, and to reproduced the signal strength $\mu_{13}^{\gamma \gamma} \approx 7 \fb$, where it is important to recall that both the signal strength and the width scale with $1/f^2$. 
For values of $C_{\gamma \gamma}$ close to the lower bound, a non-vanishing UV coupling to gluons  $C_{gg} \approx 3$ is required.
However, for larger couplings to photons, $C_{\gamma \gamma} \approx 110 \, (f/500\GeV)$, the signal strength and width can be reproduced with the effective coupling to gluons being generated only via the top loop, that is $C_{gg} = 0$.  
Instead, if the requirement of a large width is relaxed, smaller values of $C_{\gamma\gamma}$ and $C_t$ can reproduce the excess.
In fact, while $C_{\gamma\gamma} \neq 0$ is always needed, the $\eta$ coupling to tops could be completely absent, as long as $C_{gg} \neq 0$.
For $f = 500 \GeV$, values as small as $C_{\gamma\gamma} \approx 5$ and $C_{gg} \approx 0.5$ could yield the right signal strength, compatibly with dijet searches \eq{jjbound}.
Notice that if both the couplings of $\eta$ to tops and to gluons are negligible, then photon fusion becomes the dominant production mechanism, which currently appears to be in tension with 8 TeV bounds from $\gamma\gamma$ searches.\footnote{This statement is, at present, subject of discussion: 
while the ratio $\sigma_{13 \rm TeV}/\sigma_{8 \rm TeV}$ extracted from photon parton distribution functions is close to 2 (see e.g. \cite{Fichet:2015vvy,Csaki:2016raa}), and thus excluded, see the claim in \cite{Fichet:2015vvy} that the finite size of the proton could lift the same ratio to $\approx 4$.} 
These points are illustrated in Figure \ref{fig:goldstone} left-hand plot, where contours of $\Gamma_\eta$ and $C_{\gamma \gamma} \, (500\GeV/f)$ are plotted in the plane of $C_t/C_{\gamma \gamma}$ and $C_{gg}/C_{\gamma \gamma}$. 
In the right-hand plot of Figure \ref{fig:goldstone} we show contours of the expected signal strength (normalized to $\mu^{\gamma \gamma}$) for the production and decay of $\eta$ into $t \bar t$ and $gg$. These channels could provide very valuable information on the GB explanation of the diphoton anomaly.\footnote{
For an earlier study of similar signatures at the LHC of axion-like particles see e.g.~\cite{Jaeckel:2012yz}.} For example, a sensitivity to top pair production cross sections below $\sim 2$ pb (assuming $\mu_{13}^{\gamma \gamma} = 7$ fb) would probe regions of parameter space still unexplored, along with the possibility that a large width is generated by decays to gauge bosons and $t \bar t$ alone.

In the discussion above we have omitted the couplings of the GB to $ZZ$, $Z\gamma$, and $W^+ W^-$, all of them linked to the coupling to $\gamma \gamma$, see \eq{looprelations}. 
We have done so for simplicity of the exposition, to focus on the minimal ingredients required for the GB to reproduce the diphoton excesss. 
Indeed, one can explicitly check that if the GB's anomalous terms are such that $C_B \gg C_W$, the constrains from 8 TeV searches into these other channels are easily evaded (notice the $\tan \theta_W$ suppression of $C_{ZZ}$ and $C_{Z \gamma}$ compared to $C_{\gamma \gamma}$ in this limit). Figure \ref{fig:goldstone} has been drawn in this limit, \ie setting $C_W = 0$ in the relations \eq{looprelations}, and neglecting the top loop contribution to the $ZZ, Z\gamma, WW$ decays, since this becomes important for our discussion only in a region where $C_t$ is very suppressed.
This however should not be taken as a solid feature of the model, on the contrary 13 TeV searches of $gg \to \eta \to ZZ, Z\gamma, W^+ W^-$ should be an important target for future experimental analyses. \\

Finally, let us discuss the origin and the expected size of the pGB anomaly terms and mass. 
The size of the anomaly coefficients needed to reproduce the diphoton excess indicates that there should be a relatively large number of UV degrees of freedom charged under the SM gauge groups contributing to the anomalies. 
This is actually plausible in large-$N$ theories with non-trivial colour and electromagnetic charges, such as composite Higgs models with partial compositeness \cite{Kaplan:1991dc} and their holographic realizations \cite{ArkaniHamed:2000ds,Contino:2004vy,Agashe:2004rs}. 
In such kind of models, besides the aforementioned contributions from UV constituents, light IR d.o.f.'s like the composite top partners could contribute significantly at one-loop to the $\eta F_{\mu \nu} \tilde F^{\mu \nu}$ effective couplings. 
However, one should recall that this type of operators explicitly break the $U(1)_\eta$ shift symmetry, and since $\eta$ is SM neutral, extra couplings breaking explicitly the symmetry, such as $C_t y_t$, $y_t = m_t\sqrt{2}/v$, must be involved in the generation of the loop-induced $\eta$ interactions \cite{Giudice:2007fh}. 
Therefore the GB nature of $\eta$ implies that the IR contributions scale as $\Delta C_{\gamma \gamma} \sim C_t y_t  f/m_*$, where $m_*$ is the mass of the composites. 
Most likely these contributions are too small to account for the $C_{\gamma \gamma}$ coupling required to explain the diphoton signal (given $m_* \gtrsim f$), although they could gives rise to the required $C_{gg}$ coupling in some regions of the parameter space. 
Importantly, this is in contrast with the contributions from vector-like fermions to the couplings of a generic (pseudo-)scalar particle. 
In such a case loops of enough states with sufficiently large SM charges could give rise to the required $gg$ and $\gamma \gamma$ couplings, but at the expense of uncontrollable contributions to the mass of the diphoton candidate, generically pushing it to the cutoff of the effective theory \cite{Low:2009di}. 
In this respect, the mass of the GB is controlled by the interactions that explicitly break the global $U(1)_\eta$ symmetry. 
In our minimal case we have two candidate sources: the anomalies and the top Yukawa. 
Other possible breaking terms can be taken naturally small, being controlled by a symmetry. 
The anomaly terms, even though they may come with sizeable coefficients, are totally harmless and contribute very little, if at all, to the mass of the pGB. 
Indeed, the anomalous coupling to photons is invariant under the shift symmetry $\eta\rightarrow \eta+c$ up to a boundary term that can be dropped given the trivial topological structure of $U(1)_{em}$. 
The anomalous coupling to gluons would seem more problematic as the action admits topologically non-trivial stationary configurations that vanish only as $1/r$, aka instantons, for which the boundary integral gives a finite contribution. 
However, since the relevant scale in this case is $\Lambda_{\mathrm{QCD}}$, which is smaller than $f$ by several orders of magnitudes, the effect on the $\eta$ mass is totally negligible scaling as $\Delta m^{2}_\eta \sim m_u \Lambda_{\mathrm{QCD}}^3/f^2$. This insensitivity to sizeable anomalies for the pGB should be contrasted with the case of the dilaton discussed in Section~\ref{sectDil}, see Eq.~(\ref{dilatonmassestimate}). 
Whether the Yukawa terms in \eq{yuketa} contribute or not to the mass is a model dependent assumption which is reflected on the non-linear terms in $\eta$, which we have not specified.  For example, should the Yukawa of the top quark be completed by $-e^{i C_t \eta/f} y_t \bar{q}_L \tilde H t_R $, we could rotate away $\eta$ with a field redefinition of the type (\ref{Qeta}) at the expense of generating a symmetry preserving coupling to the current \eq{etacurrent}, which has no effect on the mass, as well as the anomalous couplings to photons and gluons which has little or no effect on the mass as we have already discussed (in this case $\eta$ is essentially a heavy axion-like particle). On the other hand, should the non-linear structure for $\eta$ be such that the linear coupling of $\eta$ to the top is physical, the Yukawa would be a genuine extra source of breaking: the associated contribution to the mass, similar to the top loop contribution to the Higgs mass, can be estimated as
\beq
(\Delta m^{2}_\eta)_t \sim 3 \frac{C_t^2 y_t^2}{8 \pi^2} m_*^2 \approx (775 \GeV)^2 \left( \frac{C_t}{2} \right)^2 \left( \frac{m_*}{2 \TeV} \right)^2 \, . 
\eeq
Such a contribution disfavours $C_t \gg 1$.
Nonetheless, it is interesting that with the same value $C_t \approx 2$ both the pGB mass and large width, Eq.~(\ref{ttwidth}), can be naturally reproduced. 
This fact crucially hinges on the relatively low value $f \approx 500 \GeV$, given that $\Gamma_{t \bar t}/m_\eta \sim (v^2/f^2) (m^{2}_\eta/f^2)$. 
Notice that $v^2/f^2$ is a measure of fine-tuning in the Higgs sector, thus low values of the GB decay constant are preferred (as long as they are compatible with precision measurements).

\section{Sgoldstino}
\label{sectSgold}

We discuss now a model where $\Phi$ has a tree-level size coupling to photons and gluons. We identify $\Phi$ with the sgoldstino, the partner of the goldstino $G$ that emerges from the spontaneous breaking of $\mathcal{N}=1$ SUSY. 
For earlier studies of sgoldstino phenomenology see e.g.~\cite{Perazzi:2000id,Gorbunov:2002er,Brignole:2003cm,Petersson:2011in,Bellazzini:2012mh,Dudas:2012fa,Dudas:2013mia,Petersson:2015rza} and references therein. 
The sgoldstino and the goldstino live in a chiral superfield $X=\Phi+\sqrt{2}\theta G+\theta^2 F_X$, which is a gauge singlet that gets VEV in the auxiliary field component $\langle F_X\rangle=\mathcal{F}$. $\Phi = (\sigma+i\eta)/\sqrt{2}$ has a CP-even and a CP-odd component, $\sigma$ and $\eta$ respectively. We assume low-scale SUSY breaking, $\sqrt{\mathcal{F}}\ll M_{\mathrm{Pl}}$ (and in fact $\sqrt{\mathcal{F}}$ in the TeV range), such that the transverse gravitino components decouple and we can treat SUSY as a global symmetry in accordance to the supersymmetric equivalence theorem \cite{fayet,Casalbuoni:1988kv,Casalbuoni:1988qd}. The sgoldstino is not protected by the Goldstone shift symmetry and gets a mass $\sim \mathcal{F}/M$ where $M$ is the characteristic scale of the heavy states that mediate SUSY breaking. However, the overall size of the coefficient is model dependent and it can arise either at loop- or tree-level (the difference being blurred in strongly coupled models).  As for every Goldstone multiplet, the interactions originate from the coupling to the conserved (super)current. Equivalently, the couplings can be read off the soft breaking terms of SUSY, which are compensated by the would-be spurion $X$ which has been promoted to a dynamical field.
For example, the gaugino masses $m_i$ as well as the sgoldstino couplings to gauge bosons come from 
\begin{equation}
-\frac{1}{2\mathcal{F}} \int d^2\theta X\left(m_1 W^{\alpha}  W_\alpha + m_2 W^{\alpha\, a_2}  W^{a_2}_\alpha +m_3 W^{\alpha\, a_3}  W^{a_3}_\alpha\right) + h.c. \, .
\end{equation}
Analogously,  the top $A$-term gives the stop mixing mass and a Yukawa coupling to $\Phi$: 
\begin{equation}
-\frac{A_t}{\mathcal{F}}\int d^2\theta X Q_L H_u t_R + h.c. \, .
\end{equation}
With the $A$-terms proportional to the Higgs Yukawa couplings, only $A_t$ is potentially relevant.  For concreteness, we focus hereafter on the phenomenology of the CP-even component of the sgoldstino, $\Phi=\sigma$. Picking then the $\sigma$ component of $X$, we generate possibly large couplings to the transverse gauge bosons which, in terms of the parametrization of Eq.~(\ref{ffsigma}), read
\begin{align}
\frac{c_{gg}}{f} = \frac{2\sqrt{2}\pi m_3}{\alpha_3 \mathcal{F}} \, , & \qquad \frac{c_{\gamma\gamma}}{f}=  \frac{2\sqrt{2}\pi m_{\sigma\gamma\gamma}}{\alpha_e \mathcal{F}} \, , \\
 \frac{c_{WW}}{f}= \frac{2\sqrt{2}\pi m_2}{\alpha_2 \mathcal{F}} \, , & \qquad \frac{c_{ZZ}}{f}=  \frac{2\sqrt{2}\pi m_{\sigma ZZ}}{\alpha_2\cos^2\theta_W \mathcal{F}} \, ,
\qquad \frac{c_{Z\gamma}}{f}=  \frac{2\sqrt{2}\pi  \tan\theta_W m_{\sigma Z\gamma}}{\alpha_e \mathcal{F}} \, ,
\end{align}
where $m_{\sigma\gamma\gamma}=m_1\cos^2\theta_W + m_2 \sin^2\theta_W$, $m_{\sigma ZZ}=m_1\sin^2\theta_W + m_2 \cos^2\theta_W$,  and $m_{\sigma Z\gamma}=(m_2-m_1)\sin\theta_W \cos\theta_W$. Notice that all these effective couplings scale with the ratios $m_i/\mathcal{F}$. 
The partial widths into gauge bosons, at leading order in $m_V/m_\sigma$, then read 
 \begin{align}
 \Gamma_{gg}= \left(\frac{m_3}{2 \mathcal{F}} \right)^2 \frac{m_{\sigma}^3}{\pi}\,, & \qquad
 \Gamma_{\gamma\gamma}= \frac{1}{2} \left(\frac{m_{\sigma\gamma\gamma}}{4 \mathcal{F}} \right)^2 \frac{m_{\sigma}^3}{\pi}\,, \\ 
 \Gamma_{ZZ} \simeq \frac{1}{2} \left(\frac{m_{\sigma ZZ}}{4 \mathcal{F}} \right)^2 \frac{m_{\sigma}^3}{\pi}\,, & \qquad
 \Gamma_{WW} \simeq \left(\frac{m_{2}}{4 \mathcal{F}} \right)^2 \frac{m_{\sigma}^3}{\pi}\,, \qquad 
 \Gamma_{Z\gamma} \simeq \left(\frac{m_{\sigma Z\gamma}}{4 \mathcal{F}} \right)^2 \frac{m_{\sigma}^3}{\pi}\,. 
 \end{align}
Other subleading contributions come from the coupling to longitudinal vector bosons, $\sigma V_\mu V^\mu$, see for instance \cite{Bellazzini:2012mh,Perazzi:2000id}. A contribution to such interactions comes from a mixing of $\sigma$ with the Higgs boson, controlled e.g.~by the $\mu$-term. However for $\mu/m_{1,2} \lesssim 1/2$ the corrections to the partial widths above are below the experimental sensitivity. As a matter of fact, to leading order in the gauge boson masses, the expressions above are still valid with the replacement $m_{\sigma ZZ}^{2} \to m_{\sigma ZZ}^{2} + \mu^{2}/2$ for $\Gamma_{ZZ}$ (and $m_2^2\rightarrow m_2^2+\mu^2/2$ for $\Gamma_{WW}$ as well).  The $\mu$-term in these replacements is actually the mass entry of the neutral Higgsinos, $\mu \tilde{H}^0_1 \tilde{H}^0_2$, which is part of the full neutralino mass matrix that can be found in \cite{Brignole:2003cm}.  A sizeable mixing with the Higgs may be problematic to explain the supposed signal, given the bounds at 8 TeV, in particular those from $ZZ$ and $W^+W^-$ resonance searches. We work here in the limit where this mixing is negligible, an assumption justified also by measurements of the Higgs signal strengths, in particular in the diphoton channel. 
The widths translate into the following ratios of decay rates
 \begin{align}
 r^\gamma_{gg} = \frac{1}{8}\frac{m^2_{\sigma\gamma\gamma}}{m_3^2}\,,\qquad 
 r^\gamma_{ZZ} \simeq \frac{m^2_{\sigma\gamma\gamma}}{m_{\sigma ZZ}^2}\,,\qquad 
 r^\gamma_{Z\gamma} \simeq \frac{1}{2}\frac{m^2_{\sigma\gamma\gamma}}{m_{\sigma Z\gamma}^2}\,,\qquad 
  r^\gamma_{WW} \simeq \frac{1}{2}\frac{m^2_{\sigma\gamma\gamma}}{m_{2}^2}\,,
 \end{align}
 which are a priori well suited to fit the diphoton excess consistently with 8 TeV searches. From these ratios the constraints on $r^\gamma_{VV}$ and $r^\gamma_{gg}$ in Table~\ref{tab2}  can be easily satisfied in a region of parameter space where $m_{3}/m_{1}$ and $m_{2}/m_{1}$ are not too large, as shown in Figure~\ref{fig:sgoldstino}. 
It is important to notice that the sgoldstino mass term implies an irreducible invisible width because of the decay channel into goldstinos (aka the longitudinal polarizations of the gravitino after coupling SUSY to gravity \cite{fayet})
 \begin{equation}
 \Gamma_{\mathrm{inv}} \simeq \frac{m_\sigma^5}{32\pi \mathcal{F}^2}\,,\qquad r^\gamma_{\mathrm{inv}} \simeq \frac{m^2_{\sigma\gamma\gamma}}{m_\sigma^2}\,,
 \end{equation}
which originates from the soft SUSY breaking term $-m_\sigma^2/(4\mathcal{F}^2) \int d^4\theta (X^\dagger X)^2$.
The constraint on $r^\gamma_{\mathrm{inv}}$ from invisible decays of $\sigma$ (which can be derived from the monojet searches of \cite{Khachatryan:2014rra}) is very mild, and automatically satisfied given the existing lower bounds on gaugino masses. As a matter of fact, this width is typically small and we have neglected it in obtaining Figure~\ref{fig:sgoldstino}. 

\begin{figure}[!t]
\centering
\includegraphics[width=8.5 cm]{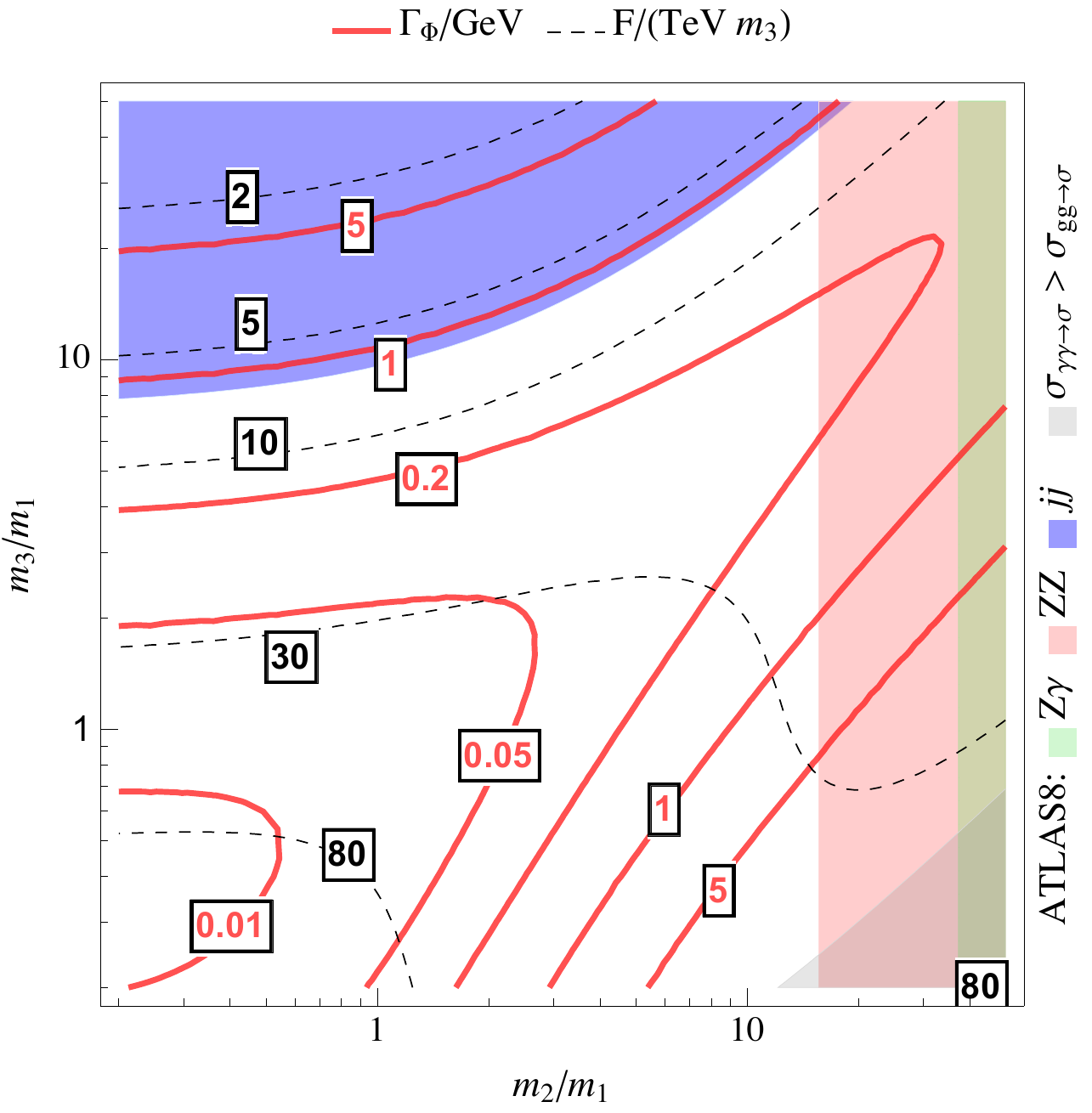}\hfill
\includegraphics[width=8.5 cm]{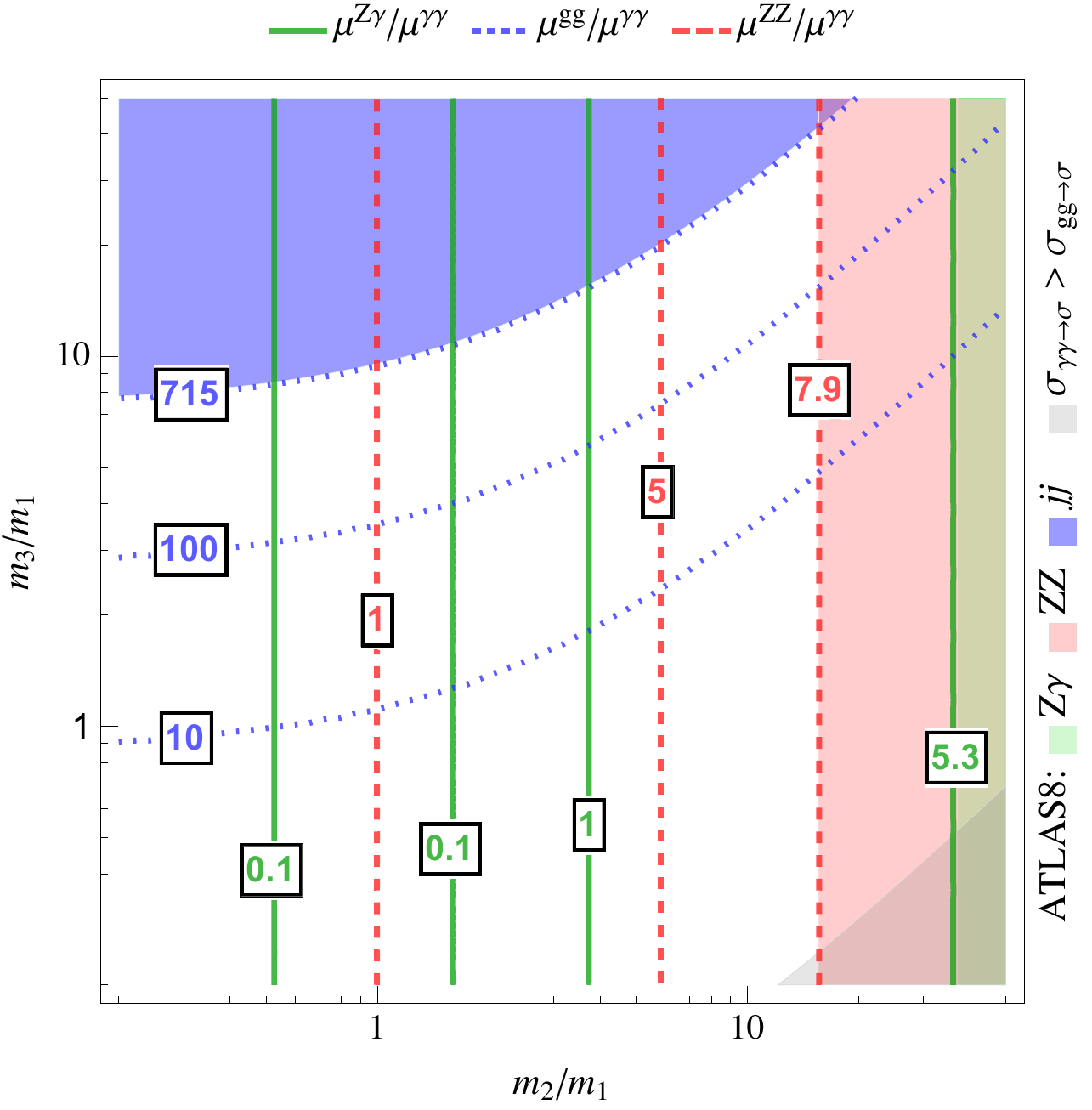}
\caption{Sgoldstino parameter space, where we have fixed $m_\sigma = 750$ GeV and $\mu^{\gamma\gamma}_{13} = 7 \fb$. Shaded regions are excluded by dijet (blue), $ZZ$ (red) and $Z\gamma$ (green) searches at 8 TeV. We have also shaded in grey the region where production via photon fusion dominates over gluon fusion (bottom-right corner). Left: the solid red line gives the expected width of the minimal sgoldstino model, the dashed black line the scale $\mathcal{F}/(\TeV \cdot m_{3})$. Right: ratio of signals in $ZZ$ (dashed red), $Z\gamma$ (solid green) and $gg$ (dotted blue) with respect to the signal in $\gamma\gamma$.}
\label{fig:sgoldstino}
\end{figure}

The production cross section of the sgoldstino scales proportionally to $(m_{3}/\mathcal{F})^2$, and the signal strength can be written as 
\begin{equation}
\mu^{\gamma\gamma}_{13,\, \sigma} \simeq 7\,{\rm fb} \left(\frac{14\TeV}{\sqrt{\mathcal{F}}}\right)^4 \left(\frac{m_3}{3\TeV}\right)^2
\frac{1}{(m_3/m_2)^2 + 1/2 + 0.07 (\TeV/m_2)^2} \, ,
\end{equation}
where for simplicity in this formula we have assumed $m_1=m_2$. In the left-hand plot of Figure~\ref{fig:sgoldstino} we show isolines (dashed black) for the values of $\mathcal{F}/(\TeV\cdot m_{3})$ that correspond to signal strength that best fits the excess. Dijet searches at the LHC Run-1 require $\mathcal{F}/(\TeV \cdot m_{3}) \gtrsim 6$, which translates into a lower bound on the gluino mass. Such a bound may or may not be relevant depending on the specific SUSY breaking model.
Notice that $ZZ$ searches at 8 TeV exclude the region where photon fusion (plus possibly $ZZ$ and $Z\gamma$ production) dominates over gluon fusion.
The sgoldstino total width can be written as
\begin{equation}
\Gamma_\sigma \simeq 8 \MeV \left(\frac{14\TeV}{\sqrt{\mathcal{F}}}\right)^4 \left(\frac{m_3}{3\TeV}\right)^2
\left[1 + {1 \over 2} (m_2/m_3)^2+ 0.07 (\TeV/m_3)^2 \right] \, ,
\end{equation}
so that the generic expectation is that the minimal sgolstino appears as a narrow resonance. Contours of fixed widths, again with the assumption of negligible invisible width, are shown as thick red lines in the left-hand plot of Figure~\ref{fig:sgoldstino}. 
These values are too small compared with the preferred width determination by ATLAS.  However, given the little amount of data currently available, we do not consider this as a major drawback of the sgoldstino model, and we postpone a more careful analysis of the total width to when more data is available (including an error for the measurement of the width).
Likewise, we leave for future work the exploration of the parameter regions of the scalar potential and/or non-minimal models with an extended Higgs sector beyond the MSSM, where it could be possible to generate a larger width in invisible or other channels, resulting in a broader resonance. 
Finally, in the right-hand plot of Figure~\ref{fig:sgoldstino}, we show the signal expected in the most promising channels to confirm the hypothesis of the sgoldstino, if this is indeed responsible for the 750 GeV diphoton excess. This plot can provide a useful guidance to experimental collaborations, providing sensitivity targets for dijet, $ZZ$ and $Z\gamma$ resonance searches. For example, a sensitivity to cross sections in $Z\gamma$ below $\sim 20$ fb (assuming $\mu_{13}^{\gamma\gamma} = 7$ fb) would probe regions of parameter space that are still unexplored, and constrain the gaugino spectrum.

While we focussed on the CP-scalar $\sigma$ in $X$, its CP-odd component $\eta$ has very similar coupling structures, dictated as well by the soft SUSY breaking terms. The phenomenology of such pseudo-scalar is fully analogous to the analysis we presented above for $\sigma$.

\section{Dilaton} \label{sectDil}

In the presence of large diphoton rates, the dilaton, \ie the GB of spontaneously broken conformal symmetry, is perhaps one of the first scalar resonances that comes to mind in the context of strongly coupled models.\footnote{We should remark that explicit constructions of spontaneous breaking of conformality, without SUSY, require strong dynamical assumptions \cite{Rattazzi:2000hs,Chacko:2012sy,Bellazzini:2012vz}, see \cite{Bellazzini:2013fga,Coradeschi:2013gda,Megias:2014iwa,Cox:2014zea,Cleary:2015pva} for explicit realizations.}
Indeed, the dilaton has potentially sizeable couplings to massless gauge bosons, controlled by the $\beta$-function contributions from the states of the conformal field theory (CFT), see e.g.~\cite{Csaki:2007ns,Goldberger:2008zz,Vecchi:2010gj,Barger:2011nu,Chacko:2012vm,Chacko:2012sy,Bellazzini:2012vz,Cox:2013rva,Chacko:2014pqa}. 
We consider a strongly coupled CFT with a global $SU(3)_C \times SU(2)_L \times U(1)_Y$ symmetry, whose associated conserved currents weakly couple to gluons and electroweak gauge bosons. 
The current central charges $\kappa_J$ control the $\beta$-functions contributions  
\begin{equation}
\label{defcentralc}
\langle J^a_\mu(p) J^b_\nu(-p) \rangle_{\mathrm{CFT}}=\delta^{ab}\left(\eta_{\mu\nu}p^2- p_{\mu} p_{\nu}\right) \frac{\kappa_{J}}{16\pi^2} \log p^2+ \ldots\,, \qquad \beta_J^{\mathrm{CFT}}=\kappa_J \frac{g_J^3}{16\pi^2}\,,
\end{equation}
where $g_J$ is the relevant gauge coupling between the gauge field and its current $J_\mu$.  The central charge roughly counts how many degrees of freedom in the CFT are charged under the symmetry. Spontaneous breaking of conformality generates a light dilaton $\sigma$ with its decay constant $f$.
The compensator $\sigma/f$ must match in the IR the trace anomaly contribution from the CFT triggered by weak gauging in the UV
\begin{equation}
\label{tracecftsigma}
\frac{\sigma}{f} T^{\mu\, \mathrm{CFT}}_\mu= \frac{\sigma}{f} \left( \frac{\alpha_3}{8\pi} \kappa_3 G_{\mu\nu}^2 + \kappa_{e} \frac{\alpha_e}{8\pi} F^2_{\mu\nu} -y_{t}(1+\epsilon_t)\bar{q}_L \tilde H t_R +h.c. +2(1+\epsilon_H) |D_\mu H|^2 +\ldots \right) \, .
\end{equation}
We have also included in the trace the contributions coming from the terms responsible for generating the masses of matter fields and electroweak gauge boson, where the electroweak symmetry is spontaneously broken by a Higgs operator $H$, which should be part of the CFT in order not to introduce a hierarchy problem. 
Within partial compositeness \cite{Kaplan:1991dc}, elementary fermions $\psi_{L,R}$ probe the CFT by mixing in the UV with some fermionic operators $\Psi^{CFT}_{R,L}$ of scaling dimension $\Delta_{L,R}=5/2+\gamma_{L,R}$, with the strength of the interaction set by the proto-Yukawas $y_{L,R}$. The resulting IR Yukawa coupling for the massless SM fermion emerging from the mixing scales as $y_{f}\sim y_L y_R$. 
Dilaton insertions $e^{\Delta_f \sigma/f}$ must then compensate in the IR a dimension $\Delta_{f}=1+\epsilon_{f}$, where $\epsilon_{f}=\gamma_{L}+\gamma_{R}$. 
We will assume in the following that for the case of the top quark $\epsilon_{t}=0$, although this is not an essential (nor very relevant) feature of the model.\footnote{The fermionic anomalous dimensions $\gamma_{L,R}$ depend on the particular model of flavour, and special cases such as $\epsilon_{t} \simeq-1$ could also be realized.} 
Analogously, one must insert $e^{2\sigma/f}$ to compensate for the gauge bosons mass terms, of conformal weight equal to two, 
with possible departures parametrized by $\epsilon_H$.
We have further assumed that electroweak symmetry breaking proceeds while respecting custodial symmetry, which enforces $c_W=c_Z=c_h$ in \eq{ffsigma}. 
Moreover, in composite Higgs models where $H$ arises as a pseudo-Goldstone boson of the strong dynamics one obtains $\epsilon_H \simeq 0$, since such a Higgs is fully composite and a possible Higgs-dilaton kinetic mixing is suppressed by the Higgs' shift symmetry \cite{Giudice:2000av,Csaki:2000zn} (we come back to this point below).
All in all, for the minimal and natural dilaton model we get  
\begin{equation}
\label{choiceCs}
c_{gg}=\kappa_3\,,\qquad c_{\gamma\gamma}=\kappa_e\,,\qquad c_W=c_Z=c_h\simeq c_t\simeq 1\, ,
\end{equation}
for the parameters in the effective lagrangian in Eq.~(\ref{ffsigma}). 
The effective couplings to photons and gluons are then given by
\begin{equation}
\label{Ceffdilaton}
c^{eff}_{\gamma\gamma}=c_{\gamma\gamma}+ c_t N_c^{(t)} Q_t^2 F_{1/2}(x_t)  -  c_W F_1(x_W)   \,,\qquad c^{eff}_{gg}=c_{gg}+ c_t \frac{1}{2} F_{1/2}(x_t)
\end{equation}
where $N_c^{(t)}Q_t^2=3\times\left(2/3\right)^2$,  $x_i = 4 m_i^2/m_\sigma^2$, and $F_{1/2}(x) =  2 x [1+(1-x) g(x)]$, $F_{1}(x) = 2 + 3 x + 3 x (2-x) g(x)$ with $g(x) = [\sin^{-1}(1/\sqrt{x})]^2$. 
The coefficients in \eq{Ceffdilaton} set the decay rates for $\sigma \to \gamma \gamma$ and $\sigma \to gg$ as in the pGB case, \eq{Gammaeta}, with $m_\eta \to m_\sigma$ and $C_{gg,\gamma\gamma}^{eff} \to c_{gg,\gamma\gamma}^{eff}/2$. 
The unknown UV contributions to $c_{gg,\gamma\gamma}^{eff}$,  parametrized by $\kappa_{3}$ and $\kappa_e$, allow us to treat these coefficients as free parameters in this model. Actually, the large rates in diphotons needed to reproduce the supposed excess, consistently with 8 TeV searches, call for values of $\kappa_{e, 3}$ that dominate the expressions for $c^{eff}_{\gamma\gamma,gg}$. Hence in the following we will focus on the implications of the diphoton signal for the $\beta$-function coefficients defined in Eq.~(\ref{defcentralc}).

A significant challenge for the dilaton scenario of \eq{choiceCs} is posed by the predicted large rate in $ZZ$,  $W^+W^-$ and $hh$,
\beq
\Gamma_{ZZ} \simeq \Gamma_{WW}/2 \simeq \Gamma_{hh} \simeq \frac{m_\sigma^3}{32 \pi f^2} \, ,
\eeq
given that the corresponding couplings are generated at tree-level, while the coupling to photons is a priori of loop-size. 
This implies, for the most sensitive channel at the LHC Run-1,
\begin{equation}
r_{ZZ}^{\gamma} \approx 0.05 \left( \frac{\kappa_e}{240}\right)^{2}\,.
\end{equation}
Therefore, in order to satisfy the bounds from $ZZ$ final state searches in Table \ref{tab2},  the contribution from the UV d.o.f.'s of the CFT to $\kappa_{e}$ needs to roughly compensate for the smallness of $\alpha_{e}=e^2/(4\pi)$. 
Considering interference effects, which could be non-negligible given the width of $\Phi$ that is suggested by the recent Run-2 diphoton data, the limit from $ZZ$ searches could not apply or be considerably loosen, hence opening the possibility to fit the diphoton excess with smaller values of $\kappa_e$. However, even in that case the combined limits from $ZZ+W^+W^-$ searches, which cover the case of new physics resonances with a large width, would apply and the lower bound on $\kappa_e$ would be reduced only by a factor less than half, leaving unchanged the conclusion that a very large electromagnetic trace anomaly is needed to attain the needed signal rates without clashing with Run-1 searches.

Such a large electromagnetic $\beta$-function is the signal of a substantial breaking of the conformal symmetry. 
This feeds back into the dilaton mass, which is expected to scale as \cite{Chacko:2012sy,Bellazzini:2013fga}
\begin{equation}
\label{dilatonmassestimate}
\frac{m_\sigma^2}{\Lambda^2} \sim \alpha_e \kappa_e
\end{equation}
where $\Lambda$ is the scale associated with the heavy resonances of the strongly coupled CFT. For large $\kappa_e \gtrsim 1/\alpha_e$ as those needed to reproduce the diphoton excess, the dilaton is then expected to be as heavy as a generic composite resonance. 
The calculability and selection rules associated with scale invariance, whose corrections scale precisely as $(m_\sigma/\Lambda)^2$, are generically lost in this case. 
The dilaton is then expected to behave as an ordinary composite spin-0 resonance of the strong sector, perhaps accidentally light, and which nevertheless could still be well suited to explain the diphoton excess. 

One tentative solution for this problem would be to depart from the tree-level couplings of the dilaton to the massive vector bosons given in \eq{choiceCs}. 
Given a modification $c_{V} \to c_{V} (1+\delta_{V})$, with $\delta_V < 0$, $V=\{W,\, Z\}$,  the ratio of $\gamma \gamma$ to $ZZ$ decay rates increases by a factor $r_{ZZ}^{\gamma} \to r_{ZZ}^{\gamma}/(1+\delta_V)^2$.
The absence of a signal at the LHC Run-1 in $ZZ$ resonance searches requires quite large deviations $\delta_{V} \gtrsim 1/3$. Retaining the assumption that the Higgs is a pGB of the strong dynamics, such a large correction indicates that the expansion parameter of explicit conformal symmetry breaking $m^{2}_{\sigma}/\Lambda^{2} \sim \delta_{V}$ should be large. 
Alternatively, one could consider a composite Higgs which is not a pGB.
In such a case, a kinetic dilaton-Higgs mixing of the type 
\begin{equation}
\label{dilhigss}
c\, H^\dagger D^\mu H \partial_\mu\sigma/f + h.c.
\end{equation}
results in $O(1)$ corrections to the dilaton coupling to vectors, $\delta_V =\delta c_V=- c$.\footnote{\label{footkineticmix} The dilaton-Higgs mixing term, alone, breaks also special conformal transformations $\delta x^\mu=2 (b\cdot x)x^\mu-b^\mu x^2+o(b^2)$, and it is thus forbidden in theories that are fully conformal as opposed to just scale invariant. Nevertheless, since the combination $\square\sigma/f+(\partial_\mu\sigma/f)^2$ transforms covariantly with conformal weight $2$, even with respect to special conformal transformations acting as $\sigma(x)\rightarrow \sigma(x(x^\prime))- f \log J^{1/4}(x(x^\prime))$ with $J(x)=1+8b\cdot x+o(b^2)$,  the operator  $-c|H|^2\left[\square\sigma/f+(\partial_\mu\sigma)^2/f^2\right]$  contains the dilaton-Higgs mixing term up to a total derivative and an extra interaction vertex that rescales the dilaton kinetic term by $(1 + c \, v^2/2f^2)$. From this conformal covariant expression for the kinetic mixing one can readily get $\delta c_V=-c$ by comparing it with the conformal covariant kinetic term $e^{2\sigma/f}|D_\mu (H e^{-\sigma/f})|^2$ responsible for the Higgs and massive gauge bosons coupling in Eq.~(\ref{tracecftsigma}), after the field redefinition $H\rightarrow H e^{\sigma/f}$.}  
The constraints from Higgs couplings measurements associated to such a mixing could be evaded, since the correction to an on-shell Higgs coupling $g$ scales as $\delta g/ g_{SM} \sim (g_{\sigma}/g_{SM})\times c\times (v/f) (m_h/m_\sigma)^2$, where $g_{\sigma}$ is the coupling of the dilaton to the same final states. 
For example, in the SM the Higgs coupling to photons is $g_{SM}=c^{h}_{\gamma\gamma} \approx -6.5$ at one-loop in the normalization of Eq.~(\ref{ffsigma}), and due to the mixing with the dilaton it would get a relative correction $\delta g/ g_{SM} \sim (v/f) \kappa_{e}/(36\times 6.5)$, possibly below the current experimental sensitivity for the values of $\kappa_{e}$ needed to reproduce the diphoton excess.  
However, we should notice that the term in (\ref{dilhigss}) breaks the Higgs's shift symmetry and therefore contributes to the Higgs potential at loop-level. 
For a composite Higgs without Goldstone protection, for which $c$ is sizable, one can estimate
\begin{equation}
\Delta m_H^2 \sim \frac{c}{16 \pi^2} \frac{\Lambda^4}{f^2}
\end{equation}
which contributes significantly to the tuning of the electroweak scale, $\Delta m_H^2/m_h^2 \sim c\,600\,f^2/v^2$ for $\Lambda \simeq 4\pi f$.

\section{Conclusions} 

The power of LHC Run-2 for producing new states beyond the SM offers an excellent opportunity to explore the TeV scale.
In this work we have investigated resonance production in the diphoton channel, motivated by the recent diphoton anomaly reported by the ATLAS and CMS collaborations. 
We focussed on Goldstone bosons, well-motivated candidates given their intrinsic connection with supersymmetry and composite Higgs models. 
Moreover, their low-energy predictions are mostly dictated by the non-linearly realized symmetries, and their soft or anomalous breakings. 
Their mass is under theoretical control and naturally below the cutoff. 

We studied three paradigmatic scenarios: the Goldstone boson of a spontaneously broken internal symmetry, focusing on a $U(1)$, which could be part of a larger symmetry breaking pattern like in composite Higgs models; the scalar partner of the goldstino in $\mathcal{N}=1$ supersymmetry, known as the sgoldstino; 
and the dilaton from a conformal field theory broken spontaneously. 

A CP-odd GB is well suited to fit the excess, given the suppressed coupling to longitudinal vector bosons and Higgs bosons due to CP invariance, and relatively large anomaly coefficients of $O(10)$. Interestingly, the production cross section and the decay rate to photons are mainly controlled by UV anomalies, which could thus potentially shed light on the UV completion of composite Higgs models. In particular, a sizable electromagnetic anomaly is unavoidable, whereas the required effective coupling to gluons could be due to a large coupling to top quarks, yielding also a large width of the CP-odd diphoton candidate.

The sgoldstino presents couplings to $\gamma \gamma$, $gg$, and $WW$, $ZZ$ and $Z\gamma$ which are all of the same order and sizable, thus being able to accommodate the excess with mild constraints on the spectrum of the gauginos. This model, in its minimal incarnation, prefers a small width, below $\sim 1$~GeV. 
Importantly, the observation of the sgoldstino at the LHC could provide information on the mechanism of supersymmetry breaking.

The minimal and natural dilaton model is instead disfavoured  by present data as it couples too strongly to the longitudinal vector bosons and to the Higgs boson, whereas the coupling to photons requires a huge trace anomaly, which in turn compromises the consistency of the model. The dilaton as a diphoton candidate could be rescued by reducing the couplings to the Higgs field, at the cost of increasing the amount of fine-tuning in the Higgs potential. 

The confirmation or dismissal of any of these scenarios is a matter of future exploration and more data from the LHC. For this reason, we have provided in Figures \ref{fig:goldstone} and \ref{fig:sgoldstino} the expected signals in channels other than $\gamma\gamma$, respectively for the pseudo-Goldstone boson ($jj$ and $tt$) and the sgoldstino ($jj$, $ZZ$ and $Z\gamma$).

\vspace{1cm}
\textbf{Note added:} 
The talks by CMS and ATLAS on the diphoton anomaly took place on Tuesday Dec.~15th \cite{15dec15}. 
Already by the end of that week a large amount of theoretical papers 
\cite{Harigaya:2015ezk,Mambrini:2015wyu,Backovic:2015fnp,Angelescu:2015uiz,Nakai:2015ptz,Knapen:2015dap,Buttazzo:2015txu,Pilaftsis:2015ycr,Franceschini:2015kwy,DiChiara:2015vdm}, 
\cite{Higaki,McDermott,Ellis,Low,Bellazzini:2015nxw,Gupta,Petersson:2015mkr,Molinaro:2015cwg}, 
\cite{Agrawal:2015dbf,Ahmed:2015uqt,Aloni:2015mxa,Bai:2015nbs,Becirevic:2015fmu,Bian:2015kjt,Cao:2015pto,Chakrabortty:2015hff,Chao:2015ttq,Cox:2015ckc,Csaki:2015vek,Curtin:2015jcv,Demidov:2015zqn,Dutta:2015wqh,Falkowski:2015swt,Fichet:2015vvy,Kobakhidze:2015ldh,Martinez:2015kmn,Matsuzaki:2015che,No:2015bsn}
appeared on arXiv.org as response to this input from the experiments. Such a tremendous research activity reflects the excitement of the community from the hints of new physics that we may have finally glimpsed at the LHC.

\section*{Acknowledgments}

We thank Dario Buttazzo and Diego Redigolo for useful discussions. B.B.~thanks Stephan Lavignac and Anibal Medina for interesting discussions. R.F.~thanks Roberto Contino for discussions. B.B.~is supported in part by the MIUR-FIRB grant RBFR12H1MW. F.S.~is supported by the European Research Council ({\sc Erc}) under the EU Seventh Framework Programme (FP7/2007-2013)/{\sc Erc} Starting Grant (agreement n.\ 278234 --- `{\sc NewDark}' project).



\end{document}